\documentclass{aa} %

\usepackage{graphicx}
\usepackage[figuresright]{rotating}
\begin{document}
\title{The GSC-II-based survey of ancient cool white dwarfs
\thanks{Based on observations made with the Italian
 Telescopio Nazionale Galileo (TNG) operated on the
 island of La Palma by the Centro Galileo Galilei of
  the INAF (Istituto Nazionale di Astrofisica) at
 the Spanish Observatorio del Roque de los Muchachos
 of the Instituto de Astrofisica de Canarias.}\fnmsep
  \thanks{Based on observations made with the William Herschel
  Telescope and Isaac Newton Telescope
operated on the island of La Palma by the Isaac Newton Group in
the Spanish Observatorio del Roque de los Muchachos of the
Instituto de Astrofisica de Canarias.} }

   \subtitle{I. The sample of spectroscopically confirmed WDs}

   \author{D. Carollo\inst{1}, B. Bucciarelli\inst{1}, S.T. Hodgkin\inst{3},
        M.G. Lattanzi\inst{1}, B. McLean\inst{2}, R. Morbidelli
          \inst{1}, R.L. Smart\inst{1}, A. Spagna\inst{1}, L. Terranegra\inst{4}
          }


  \offprints{D. Carollo}

   \institute{INAF-Osservatorio Astronomico di Torino
             I-10025 Pino Torinese, carollo@to.astro.it
         \and
             Space Telescope Science Institute, 3700 San Martin Drive
             Baltimore, MD 21218, USA
         \and
             Cambridge Astronomical Survey Unit, Institute of
             Astronomy, Madingley Road,Cambridge,CB3 0HA,UK
         \and
             INAF-Osservatorio Astronomico di Capodimonte Via
             Moiariello 16, I-80131 Napoli, Italy}


   \date{Received ..., Accepted}

   \abstract{
   The GSC-II white dwarf survey was designed
   to identify faint and high proper motion objects, which we used to
   define a new and independent sample of cool white dwarfs.
   With this survey we aim to derive new constraints on the halo white dwarf space density.
   Also, these data can provide information on the age of
   thick disk and halo through the analysis of the luminosity
   function. On the basis of astrometric and photometric
   parameters,
   we selected candidates with $\mu \ga 0.28
   ''yr^{-1}$ and $R_F \ga 16$ in an area of 1150 square degrees.
   Then, we separated white dwarfs
   from late type dwarfs and subdwarfs by means of the reduced proper
   motion diagram. Finally, spectroscopic
   follow-up observations were carried out to confirm the
   white dwarf nature of the selected candidates.
   We found 41 white dwarfs of which 24 are new discoveries.
   Here we present the full sample and for each object
   provide positions, absolute proper motions, photometry, and
   spectroscopy.
}
\authorrunning{Carollo et al.}
\titlerunning{The GSC-II based survey of ancient cool white dwarfs}
\maketitle
%

\section{Introduction}

Understanding the nature of the Baryonic Dark Matter in the
galactic halo is a very fundamental subject of astrophysics. The
most favoured candidates are brown dwarfs, planets, ancient cool
white dwarfs (WDs), neutron stars and primordial black holes. All
these objects are known as MACHOs (Massive Compact Halo Objects)
and, in principle, could be detected by their gravitational
microlensing effect on field stars (Pacynski 1986). The two major
experiments for the detection of microlensing events are MACHO
(Alcock et al. 2000) and EROS (Afonso et al. 2003). They measured
several events toward the Galactic bulge and the Large (LMC) and
Small Magellanic Clouds (SMC). The excess of events towards the
Magellanic Clouds is directly related to the dark matter in the
halo of our Galaxy, and their interpretation is still under
discussion. One potential explanation for these events is that the
lenses belong to the halo and, in this case, the MACHO experiment
seems to indicate that $\sim 20\%$ of the dark halo is made of
compact objects with mass of $0.5M_{\odot}$ (Alcock et al. 2000).
At the moment this evaluation of the baryonic content of the dark
halo is controversial because, recently, the EROS-2 collaboration
claimed a new value of $\sim 3\%$, using a refined set of data
(Tisserand and Milsztajn 2005). Moreover, new photometric measures
of the microlensing event MACHO-LMC-5 confirmed that the lens is
an M5 dwarf star of $0.2M_{\odot}$ and could suggest another
explanation for the microlensing events, i.e. the lenses belong to
a previously undetected component of the disk of the Milky Way
(Nguyen et al. 2004). If the lenses belong to the halo, the
natural candidates that explain the MACHOs and EROS results are
very old cool WDs with a mean mass around $0.5M_{\odot}$, and
several surveys have been carried out in order to detect a
significant number of Population II WDs (see the review by Hansen
\& Liebert (2003) and Reid (2005)).


Here we present a new survey based on the material used for the
construction of the GSC-II (Guide Star Catalogue II) (McLean et
al. 2000). This survey was developed in order to search for faint
and high proper motion objects and to define a new sample of cool
WDs. This paper presents the catalogue of the spectroscopically
confirmed WDs. Section 2 describes the GSC-II material,
astrometry, photometry, and proper motions. In Section 3 we
present the selection criteria for the candidates, while in
Section 4 we give details of the spectroscopic observations.
Finally in Section 5 we describe the WD sample and discuss some
interesting objects.

\section{The GSC-II plate material and processing}

The Guide Star Catalog II is an all-sky database based on the
photographic surveys carried out with the Schmidt telescopes of
the Palomar and Anglo-Australian (AAO) observatories. All
$6.4^{\circ}\times6.4^{\circ}$ plates were digitized at the STScI
(Space Telescope Science Institute) utilizing modified PDS type
scanning machines (15 $\mu$m/pixel, i.e. 1 arcsec/pixel). This
GSC-II database, from which the public GSC2.2 catalogue was
exported, contains multi-epoch positions, multi-band magnitudes,
and classification for $\sim$ 1 billion objects to $B_J \simeq$
21.5 mag derived from the analysis of $\sim$ 8000 plates (McLean
et al. 2000). Table 1 shows the main characteristics of the
material utilized for the construction of the GSC-II.  For each
survey we report the observation epochs, the photographic emulsion
and filter that define the passband, and the magnitude limit. For
our survey we used 32 regions of the Northern hemisphere ${\mathbf
(\delta \ga 0^\circ)}$, selected at high galactic latitudes in
order to avoid crowding, and concentrated towards the North
Galactic Pole (NGP). Figure 1 shows the distribution of the 32
regions. The total area covered is about 1150 square degrees. For
each region we used three second epoch plates (POSS-II blue, red
and infrared) and two first epoch plates (POSS-I blue and red),
and occasionally the Quick V and ER plates. To detect high proper
motion objects, we chose regions having overlapping POSS-II plates
with epoch difference of $\Delta$t $\sim$ 1-10 yr. Most of these
plates were reprocessed with the same pipeline used for the
construction of the GSC-II, in order to take advantage of the
latest photometric calibration available to the project.
In other cases, the data were extracted directly from the GSC-II
database (e.g. the POSS-I plates used to improve the precision of
the proper motions). Then, the object matching and proper motion
evaluation was performed using the procedure described in Spagna
et al. (1996).


\subsection{Astrometric and photometric calibrations}

Object positions were computed by means of astrometric
calibrations based on the reference stars extracted from the
Tycho-2 catalogue (Hog et al 2000). A quadratic polynomial was
adopted to model the transformation from the plate coordinates
x,y, previously corrected for refraction, to the standard
coordinates from which object positions $\alpha, \delta$ were
derived. Also, the typical distorsions which affect Schmidt plates
were corrected by means of astrometric masks which model the 2D
pattern of the position residuals with respect to the reference
catalog as derived from the average of about 100 plates. The final
accuracy of the absolute astrometry is better than $0.''2$ -
$0.''3$, while relative astrometry has an accuracy of $0''.1$. The
photometric calibrators used for GSC-II are the GSPC-II, GSPC-I
and Tycho catalogs, which sample, respectively, the faint, bright,
and very bright  range of the GSC-II photographic magnitudes
(Bucciarelli et al. 2001). A photometric accuracy of 0.1-0.2 mag
is generally attained. Larger errors may be present for faint
objects close to the magnitude limit of the plate.



\begin{figure}

\vspace{9cm} \includegraphics{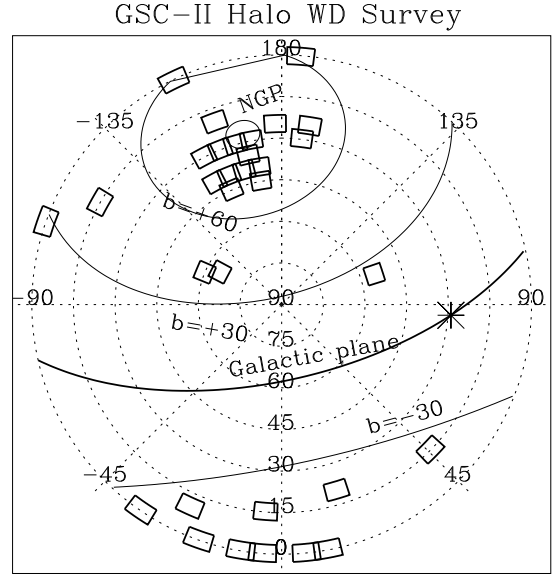}


\caption{Distribution of the GSC-II fields (boxes) in equatorial
coordinates. Lines show the location of the NGP, the galactic
plane and the strips at $b = \pm 30^\circ$ and $b = +60^\circ$. }

\end{figure}

\begin{figure}

\vspace{9cm} \includegraphics{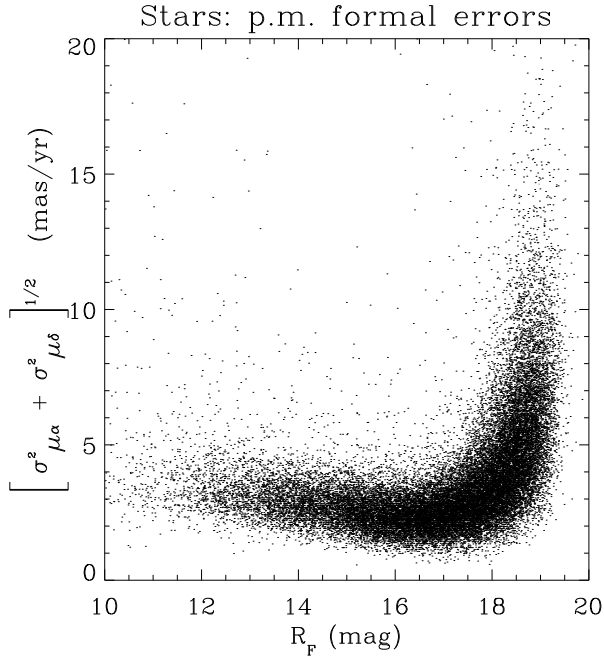}

\caption{Final proper motion errors as a function of magnitude for
one of the field towards the NGP included in this study.}

\end{figure}

\begin{table}
\caption{Main characteristics of the plate material utilized for
the construction of the GSC-II. }
\begin{tabular}{lcclllllr}

\hline\hline

  Survey&Epoch&Emulsion&Band&Depth \\
            &     &+Filter &    &     & \\
\hline
    Pal-QV    & 1983-85 & IIaD+W12    & $V_{12}$  & 19.5\\
    SERC-J    & 1975-87 & IIIaJ+GG395 & $B_{J}$ & 23.0   \\
    SERC-EJ   & 1979-88 & IIIaJ+GG395 & $B_{J}$ & 23.0   \\
    POSS-I E  & 1950-58 & 103aE+red   &   E  & 20.0   \\
    POSS-I O  & 1950-58 & 103aO       &   O  & 21.0   \\
    POSS-II J & 1987-00 & IIIaJ+GG385 & $B_{J}$ & 22.5   \\
    POSS-II F & 1987-99 & IIIaF+RG610 & $R_{F}$  & 20.8   \\
    POSS-II N & 1989-02 & IV-N +RG9   & $I_{N}$  & 19.5   \\
    AAO-SES   & 1990-00 & IIIaF+OG590 & $R_{F}$  & 22.0   \\
    SERC-ER   & 1990-98 & IIIaF+OG590 & $R_{F}$  & 22.0   \\
    SERC-I    & 1990-02 & IV-N +RG715 & $I_{N}$  & 19.5   \\
    M W Atlas & 1978-85 & IV-N +RG715 & $I_{N}$  & 19     \\
    SERC QV   & 1987-88 & IIaD +GG495 & $V_{495}$  & 14   \\
    AAO SR    & 1996-99 & IIIaF+OG590 & $R_{F}$  & 20     \\

\hline
\end{tabular}
\end{table}

\subsection{Proper motions}

Initially, relative proper motions were derived from POSS-II
plates only (epoch difference between 1-10 years) by applying the
procedure described in Spagna et al. (1996). In practice, faint
field stars ($R_F\simeq $17-18) have been selected and used as
reference stars to fit a third order polynomial and transform the
instrumental coordinates $x,y$ from the secondary plates to the
reference plate, usually the one at intermediate epoch. Then,
relative proper motions $\mu_x, \mu_y$ were computed from the
multi-epoch positions by means of a linear fit. Given a position
accuracy of $\sigma_{x,y}\simeq 0.1''$, the POSS-II plates provide
proper motions with a typical accuracy, $\sigma_{\mu_{x,y}} \simeq
\sigma_x / \Delta T$, of a few hundredths of arcsec per years
which is sufficient for the identification of high proper motion
objects and the selection of our targets. For the candidates
confirmed by the spectroscopic follow-up (Table 2) final and more
accurate proper motions were computed by using multi-epoch
positions derived from a larger set of plates including the first
epoch POSS-I, the intermediate epoch Quick-V, the second epoch
POSS-II and ER (for the equatorial fields) (cfr.\ Tab.\ 1). In
order to cross-match the objects on all these plates, the short
baseline proper motions previously computed were used to predict
their position at the different epochs. Thanks to the $\sim 40$
years
between POSS-I and POSS-II plates, the typical precision of the
final proper motions is $\sigma_\mu\sim 3$ mas~yr$^{-1}$ down to
$R_F\simeq 18$.
Finally, absolute proper motions were derived by forcing the
extragalactic sources to have a null tangential motion.

As an external check we compared the final proper motions of the
targets in Table 2 with values from USNO-B. The measurements of
the two catalogues appear well correlated, a part few outliers
possibly due to wrong USNO-B solutions.
  The rms on the total
proper motion difference is about $\sigma_{\Delta\mu} \simeq 8 $
mas~yr$^{-1}$, after rejecting 7 objects with large proper motion
difference ($|\Delta\mu_{\alpha,\delta}|>100$ mas~yr$^{-1}$) and
large internal errors
($\sqrt{\sigma_\mu{_\alpha}^2+\sigma_\mu{_\delta}^2}>20$
mas~yr$^{-1}$).


\section{Selection of the candidates}
Cool halo WDs are not easy to observe because they are quite
faint. In fact, theoretical cooling tracks by Chabrier et al.
(2000) for hydrogen atmosphere WDs predict an absolute magnitude
of $M_{R} = 15.6$ and 16.6 for a $0.6M_{\odot}$ WD of 10 and 13
Gyr respectively (excluding the nuclear burning phases). Objects
of these magnitude are in reach only within a few tens of parsecs
with the GSC-II material which contains objects down to the plate
limits of 22.5, 20.8, and 19.5 mag for the blue $B_{J}$, red
$R_{F}$, and infrared $I_{N}$ plates, respectively. In total we
reprocessed 160 plates using the standard software pipeline which
provided astrometry and photometry. As mentioned in Section 2.2 we
identified high proper motion objects using the three second epoch
plates with epoch difference between 1 yr and 10 yr.

After selecting a reference plate, usually that of intermediate
epoch, objects on different plates were matched with a variable
search radius, which grew as function of the epoch difference, so
as to include objects moving up to 2.5 $''$~yr$^{-1}$. Their
relative proper motions were then derived. To minimize false
candidates, we selected only objects which were matched on all
three POSS-II plates.





\begin{figure*}
\begin{center}

\includegraphics[height=18.5cm, width=11cm,angle=+90]{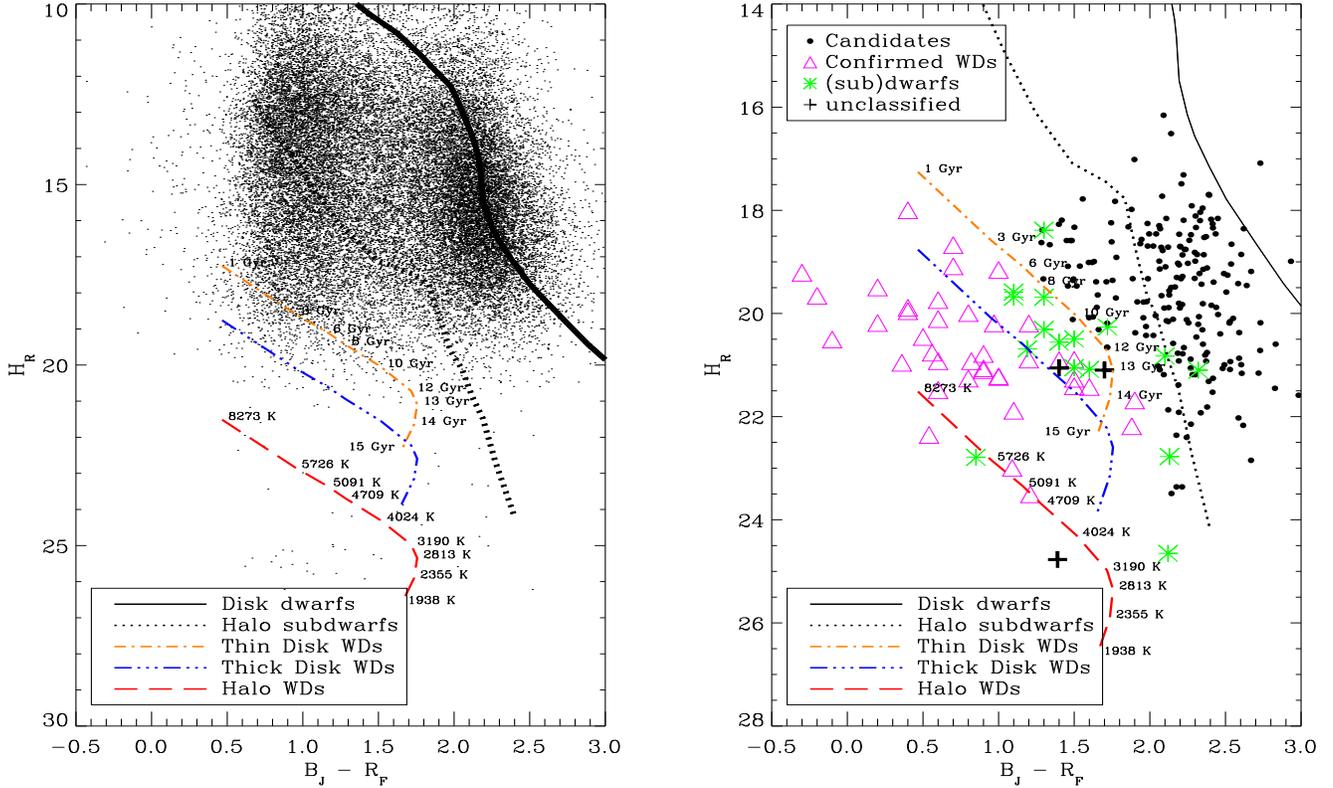}
\end{center}

\caption{Left panel: RPM diagram for a GSC-II region located
towards the NGP. Stellar isochrones and WD cooling tracks are
plotted with different lines (see text).
 Right panel: RPM diagram for all the
candidates selected in the 1150 square degrees surveyed for this
study. Dots show the unobserved targets, while triangles and
asterisks are the spectroscopically confirmed WDs and
contaminating (sub)dwarfs, respectively. Crosses indicate stars
unclassified due to the low signal to noise of their spectra.}
\end{figure*}

White dwarf candidates were selected by means of various criteria.
We searched for faint ($R_{F} \ga 16$) and fast moving stars
($0.28<\mu<2.5''$~yr$^{-1}$)\footnote{After the evaluation of the
final absolute proper motion of the confirmed WDs (see Table 2), a
few objects resulted moving with $\mu<0.28''$~yr$^{-1}$. This is
due to the relatively large error of the short baseline proper
motions (POSS-II plates only), which were used for the selection
of the candidates; in some cases these are objects close to the
plate edge. Also, improved photometry resulted in few objects
slightly brighter than $R_F=16$.}. Assuming that our survey is
efficient down to $R_F\simeq 19.0-19.5$ (i.e. 1.0-1.5 magnitude
brighter than the plate limit), we can detect ancient and faint
WDs with $M_R\simeq 16.5$ up to 30-40 pc. At this distance range,
the imposed proper motion limits select objects moving with
tangential velocities in the interval $40\la V_{\rm tan}\la 470$
km~sec$^{-1}$, which includes almost all halo WDs and a large
fraction of the thick disk WDs.


The previous criteria were implemented by means of an automatic
procedure which screens the $\sim 10^5$ objects typically
available in each field and provides a list of a few hundred high
proper motion candidates. However, not all the selected targets
had a real proper motion due to mismatches, binaries, etc. Also,
the proper motion of the faintest objects were considered
suspicious, due to the sudden increase of the proper motion error
close to the magnitude limit of the plate. For these reasons the
proper motion of each preliminary candidates was confirmed by
visual inspection of the POSS-I and POSS-II plates. Finally, a
cross-correlation with other catalogues (2MASS, LHS, NLTT, SDSS)
was also performed.

A very useful parameter to further {\it refine} target selection
and separate candidates WDs from (sub)dwarfs is the reduced proper
motion (RPM), H, defined as (Luyten 1922)
\begin{equation}
H = m + 5 \log\mu + 5 = M + 5 \log V_{tan} - 3.38
\end{equation}
where $m$ is the apparent magnitude of the star, $\mu$ ($''$/yr)
is the proper motion, $M$ is the absolute magnitude, and $V_{tan}$
(km~sec$^{-1}$) is the tangential velocity. High values of H means
faint and$/$or fast moving stars.

The reduced proper motion, used in combination with colors, is
very powerful in separating the different population of the
Galaxy.
This is evident in Figure 3 (left panel), which shows the RPM
diagram for one of the GSC-II regions located towards the NGP.
Here, the dots represent {\it all} the stars found in the field
(i.e. without any proper motion selection), and we notice that
they are clustered in two main populations. The group of objects
on the top right part of the diagram is composed of late type disk
dwarfs, while the population on the left with $B_J-R_F \approx$
1.0 is mostly subdwarfs.

White dwarfs are located on the bottom left part of the RPM
diagram\footnote{Notice that most of the objects with $H\approx
25$ are not real, and were rejected by the application of the
selection criteria and visual inspection.}. This classification is
confirmed by the position of the stellar isochrones and WD cooling
tracks shown in Figure 3. The solid and dotted lines on the top
right part of the diagram show the loci of the disk dwarfs and of
the halo subdwarfs based on the 10 Gyr isochrones down to $0.08
M_{\odot}$ of Baraffe et al. (1997, 1998) with [M/H]=0 and -1.5,
respectively. Theoretical isochrones were converted to
photographic magnitudes $B_J$ and $R_F$ using the color
transformations adopted for the photometric calibration of the
GSC-II.

The dot-dashed, dot-dot-dashed, and dashed lines are based on the
cooling tracks for $0.6 M_{\odot}$ WDs with hydrogen atmosphere
from Chabrier et al. (2000). They were obtained by adopting
tangential velocities of 38, 80, and 270 km~sec$^{-1}$, for the
thin disk, thick disk and halo, respectively.\footnote {Those
numbers represent mean values computed numerically assuming the
velocity ellipsoids $(\sigma_U, \sigma_V, \sigma_W;
v_a)=(34,21,18;-6)$ for the thin disk (Binney \& Merrifield 1998,
Tab 10.4), $(63, 39, 39; -45)$ for the thick disk (Soubiran et al.
2003), and $(160, 89, 94; -217)$ for the halo (Casertano et al.
1990). Also, a solar motion $(U,V,W)=(10.0, 5.25, 7.17)$ km
s$^{-1}$ is adopted from Dehnen \& Binney (1998).}



The shape of the cooling tracks reflects the recent improvement of
the theory of atmosphere of cool WDs. In case of a pure hydrogen
atmosphere, the cooling sequence turns back toward the blue. This
effect is due to the collision-induced-absorption (CIA) in the
infrared produced by molecular hydrogen ($H_{2}$) which gives a
redistribution of the flux toward shorter wavelength, and is
strongly present at temperatures $T_{\rm eff}<4000$ K (see, e.g.,
Chabrier et al.\ 2000).



The right panel in Figure 3 shows the reduced proper motion
diagram for all of the objects selected, in the 1150 square
degrees surveyed, after the application of the selection
procedures just described. This RPM diagram was utilized to refine
the sample of WDs candidates for spectroscopic follow-up. To
separate WDs from contaminating (sub)dwarfs, we used as boundary
line the cooling sequence for pure hydrogen atmosphere WDs (0.6
$M_\odot$) with the mean tangential velocity of the thin disk
fixed at the value in the direction of the NGP; this in practice,
corresponds to a kinematic selection of WDs faster than 38
km~s$^{-1}$. Besides a significant fraction of thin disk WDs, such
threshold includes almost all of the WDs with halo kinematics and
most of the thick disk WDs\footnote { The exact fraction of
selected objects is a function of the line of sight because it
depends on the projection of the velocity ellipsoid. Assuming the
velocity ellipsoid from Casertano et al. (1990) and Soubiran et
al. (2003) respectively for the halo and thick disk, we estimated
that for our fields at intermediate galactic latitudes less than
2\% of the halo WDs and 20\% of the thick disk WDs are moving with
$V_{\rm tan}<38$ km~s$^{-1}$. Such fraction decreases  to a
minimum of 0.25\% and 13\% of halo and thick disk WDs in the cases
of the high latitude fields towards the NGP.}.

We observed all the candidates lying below the boundary line; a
few objects above the threshold and some candidates with $B_J-R_F$
redder than the cooling track turnoff ($B_J-R_F\simeq 1.7$) were
also observed in order to check the efficiency of the selection
criteria and to take into account the photometric error.

In the figure, dots are used for the stars in the initial
candidate sample which were later excluded from the spectroscopic
follow-up list; confirmed WDs are plotted as triangles, while
(sub)dwarfs are represented as asterisks. We note that all the
confirmed WDs (except a couple of cases with slightly redder
color) fall under the boundary line, while the contaminating
(sub)dwarfs are distributed near the threshold. These results
demonstrate the good quality of the data, the selection criteria
and the accuracy of our final sample.

\section{Observations and data reduction}

Low dispersion spectra were measured to confirm the nature of the
selected candidates and remove possible contaminants, like
(sub)dwarfs.


The bulk of the spectroscopic follow-up was carried out at the
3.5m Telescopio Nazionale Galileo (TNG, La Palma), with additional
contribution of the service programmes of the 4.2m William
Herschel Telescope and the 2.4m Isaac Newton Telescope (WHT and
INT, La Palma).



We used the low resolution spectrograph DOLORES (Device Optimized
for LOw RESolution) installed at the Naysmith B focus of the TNG.
Half of our spectra were measured on the nights of April 16th-18th
2002, conditions were mostly good, the first half of the first
night suffered seeing of around 2 arcseconds, but this settled
down over the rest of the run to be 1-1.5 arcseconds. The
remaining objects were measured in a second run on October 2002,
in service time on February 2003 and in service time from August
2003 to January 2004. Most of our spectra were observed with the
low resolution, blue-optimized Grism 1 (or LR-B) with a long slit.
The transmission of the LR-B grism peaks at around 400-500 nm with
a central weavelength of 585.0 nm and useful coverage from 300 nm
to 880 nm and a dispersion of 2.8 \AA per pixel. The camera is
equipped with a 2048 $\times$ 2048 Loral thinned and
back-illuminated CCD with 15 micron pixels. The field of view is
9.38 $\times$ 9.38 arcmin with a 0.275 arcsec/pix scale, hence a 1
arcsecond slit projects to 3.6 pixels, or a resolution of $\sim$
11 \AA. Most of our targets were observed with a 1.0 arcsecond
slit; on April 15th when seeing was poor we employed the 1.5
arcsecond slit and the 2.0 arcsecond slit. Each target was
pre-imaged to define its exact position (helpful with such high
proper motion stars). Typically, an exposure time of 3600 sec and
300 sec was sufficient to provides a suitable signal-to-noise
ratio, S/N $\geq$ 10, for the faintest ($R_F \simeq$ 19.5) and the
brightest ($R_F \simeq$ 16) target respectively.
Spectrophotometric standards were observed a few times per night
to enable an approximate flux calibration, but most importantly to
help get the shape of the spectrum correct. The standards were
drawn from the list at

\begin{minipage}{2cm}
http://www.ing.iac.es/Astronomy/observing/\\
manuals/html\_manuals/tech\_notes/tn065-100/workflux.html\\
\end{minipage}

We measured the stars GD~140, BD+17~4708, Feige~67 and HZ~44. We
obtained Helium arc lamp exposures to facilitate wavelength
calibration. Halogen lamp flat fields were measured to enable us
to measure the pixel-to-pixel sensitivity variations.


Spectra for 4 objects in Tables 2-4 were measured using the
intermediate dispersion spectrographic and imaging system (ISIS)
on the 4.2m WHT. Observations were made during service nights on
the 27th and 29th January 2001, 6th December 2003, and 3rd and 4th
February 2004. ISIS is a two-arm spectrograph, employing the 5700
\AA dichroic to split the blue and red light into two cameras. In
the blue arm we used the R300B grating with the EEV12 CCD (4096
pixels in the wavelength direction) at a central wavelength of
about 4400 \AA. This setup achieves a dispersion 0.86 \AA/pix and
useful wavelength coverage of about 3000 \AA. The slit width was
matched to the seeing, typically 1 arcsecond projecting to 4
detector pixels. In the red arm we used the R316R grating with the
Marconi2 CCD (4610 pixels in the dispersion direction) at a
central wavelength of about 6250 \AA. The dispersion is 0.84
\AA/pix covering 3858 \AA, but around 1000 of these pixels suffer
significant (50-80\%) vignetting from the camera optics. Combining
the red and blue arm spectra gave us spectra with useful coverage
over the wavelength range 3400-8000 \AA.

Spectra for 2 of the objects listed in the Tables 2-4  were
measured in service time on the 2.5m INT. We employed the
(currently decommisioned) Intermediate Dispersion Spectrograph
(IDS) mounted at the Cassegrain focus. Observations were made
during the nights of 8th April 2003 and 19th  April 2003. The
instrumental setup was the same for both nights, using the 235mm
camera with the R150V grating and the EEV10 CCD. This grating and
camera combination gives a dispersion of 271.3 \AA~mm$^{-1}$,
which is equivalent to 3.66 \AA/pixel for the EEV10 detector, and
useful wavelength coverage was from approximately 3300--8700 \AA.
The camera and detector achieve a spatial scale of 0.4 arcseconds
per pixel. On the first night, the slit was opened up to 2.5
arcseconds (6 pixels). On the second night we observed with a 1.0
arcsecond slit (2.5 pixels). A total of 4 targets were measured on
the first night, with exposure times of between 1200 and 2700
seconds each, as well as one standard star, SP1550+330. On the
second night we observed 2 targets, each with 2x1800 second
exposures, as well as the flux standard SP1446+259. Calibration
arc lamps and flat fields were measured on both nights.




Data reduction for all spectra was performed within the IRAF and
MIDAS data reduction package. The same procedure was followed for
all objects wherever possible, regardless of the instrument and
telescope used to acquire the data. All images were debiased using
a mean zero second frame. Images were then flatfielded using lamp
observations to remove the pixel-to-pixel sensitivity variations.
1-D spectra were extracted and background subtracted, then
wavelength calibrated using arc lamp spectra.

Most of the spectra were flux calibrated using the standard star
observations, although no correction was made for extinction. 16
spectra have not been flux calibrated at this time and are
presented in Figure 4-6 in their uncalibrated form. Thus the
turn-over to the blue of these objects reflects the sensitivity of
the instrument, rather than the intrinsic spectral shape of the
object. This does not prevent us from identifying that these stars
are WDs, but does caution against detailed spectral analysis at
this time, which will be completed and presented in a future paper
(Carollo et al., in preparation).

\section{The GSC2 sample of WDs}

We secured spectra for 59 candidates: 40 were confirmed WDs, 16
were a mixture of late type dwarfs and subdwarfs, while the
spectra of the last 3 stars failed to reduce due to poor data
quality.
A couple of candidates below the boundary line (see Sect. 3) were
not observed for technical reasons. One of them (LHS 1093) is a
known WD (Bergeron et al. 1997). Thus, our sample counts 41 WDs in
total, 17 of which show no presence of the $H_{\alpha}$ line, and
two are exotic examples of DQ and DZ.



Table 2 gives the astrometry of the WD sample; column 1 reports
the GSC-II identification, while the remaining columns list:
positions with accuracy in the range 0.2$''$-0.3$''$, absolute
proper motions with a typical total error between 2 and 6
mas~yr$^{-1}$, and the epoch of the corresponding position.


Table 3 provides $B_{J}$ and $R_{F}$ photometry, spectroscopic
classification, and source telescope. Magnitudes are between $15.1
< R_{F} < 19.7$ and colors in the range $-0.10 <B_J-R_F <1.91$.
The last column of Table 3 lists the cross-identification with
other WD catalogues. Only 5 objects are in common with the SDSS
sample of cool WDs (Kilic et al. 2005), while 12 WDs were found in
the Villanova catalogue (McCook \& Sion 1999) available at CDS
(http://cdsweb.u-strasbg.fr/). Therefore, our sample counts 24
newly discovered WDs.

Finally, Table 4 reports the cross-match with Luyten's catalogues,
the Data Release 4 (DR4) of the SDSS (Adelman-McCarthy et al.
2005), and the 2MASS catalogue (Cutri et al. 2003); column 2 gives
the NLTT/LHS identification. In the remaining columns we provide
$ugriz$ photometry for 19 WDs found in SDSS-DR4, and $JHK_s$
magnitudes for 19 objects found in 2MASS.

Figures 4, 5, and 6 show the optical spectra of the WDs in the
GSC-II sample\footnote{The spectra of the two WDs GSC2U
J124212.6+295801 and GSC2U J222233.1+122140 have a reduced
wavelength coverage as they were observed at the TNG with a
different Grism (LR-R).}, except for the two peculiar WDs, plotted
in Figure 7. Of the 41 WDs, 17 are featureless (DC) and 19 are DA.
Two stars in the sample (LHS 5222 and LP 618-6) show carbon
features and one of them (LP 618-6) is magnetic (Schmidt et al.
2003).

Two stars (GSC2U J030339.2+140805 and LP 618-14) show the features
of a composite spectrum which indicates the presence of a red
companion. A third star (LP 618-1) is an unresolved degenerate
double formed by a magnetic DA and a massive DC (Bergeron et al.
1993).


Also, we found 2 common proper motion systems composed of WDs only
(LHS 5023 + LHS 5024 and LP 642-52 + LP 642-53). White dwarf GSC2U
J071859+551406 is the faint component of another common proper
motion pair with G 193-40 as the bright companion.

The left panel in Figure 7 shows the coolest (5120 K) carbon rich
WD known to date: GSC2U J131147.2+292348 described in Carollo et
al. (2002, 2003). Another similar DQ was recently discovered in
the SDSS (Kleinman et al. 2004); its modelling by Dufour et al.
(2005) yields a temperature of 5140 K, which increases to two the
DQs with temperature below the expected cutoff of 6000 K.\\ The
right panel shows the spectrum of GSC2U J133059.8+302955 (G 165-7)
which is a peculiar DZ with an estimated temperature of 7500 K
(Wehrse \& Liebert 1980).


\section{Conclusions}

We have presented a new high proper motion survey to search for
ancient cool WDs. The survey is based on the GSC-II material and
covers an area of about 1150 square degrees. Using appropriate
selection criteria, the RPM diagram, and spectroscopic follow-up,
we identified 41 WDs, 17 of which show no H$_\alpha$ line, and two
of peculiar nature.

This new and independent WD sample will be used to infer new
constraints on the halo WD space density (Carollo et al., in
preparation) by means of a new method to separate thick disk from
halo WDs based on the kinematic properties of the candidates
(Spagna et al. 2004). Furthermore, this sample will be useful to
increase the statistics of such rare objects and provide
information on the age of thick disk and halo through the analysis
of the luminosity function.

\acknowledgements{ The GSC-II is a joint project of the Space
Telescope Science Institute and the INAF-Osservatorio Astronomico
di Torino. Space Telescope Science Institute is operated by AURA
for NASA under contract NAS5-26555. Partial financial support to
this research cames from the Italian CNAA and the Italian Ministry
of Research (MIUR) through the COFIN-2001 program.  This work is
based on observations made with the Italian Telescopio Nazionale
Galileo (TNG) operated on the island of La Palma by the Centro
Galileo Galilei of the INAF (Istituto Nazionale di Astrofisica) at
the Spanish Observatorio del Roque de los Muchachos of the
Instituto de Astrofisica de Canarias, and also based on
observations made with the William Herschel Telescope and the
Isaac Newton Telescope operated on the island of La Palma by the
Isaac Newton Group in the Spanish Observatorio del Roque de los
Muchachos of the Instituto de Astrofisica de Canarias.\\
Special thanks to C. Loomis, A. Volpicelli and A. Zacchei for
their valuable technical support to this project and to D. Koester
for the many useful discussions. Finally, B.B. and M.G.L.
acknowledge the support of the STScI trough the Institute's
Visitor Program for 2005.}

\begin{table*}

\caption{The sample of cool WDs from the GSC-II survey:
Astrometric parameters.}


\centering                          
\begin{tabular}{llcccccc}        

\hline\hline
 ID         & RA          &  Dec          &   $\mu_{\alpha}\cos\delta$  & $\mu_{\delta}$ & Epoch\\

  ~         &(J2000)      & (J2000)       &   mas$ \cdot $yr$^{-1}$ & mas$\cdot $yr$^{-1}$ \\
\hline

GSC2U J003209.6$-$025402 &  00 32 09.57 &  -02 54 02.6  &  $630.4\pm4.4^a$ & -$149.5\pm4.3^a$ &1993.85     \\
GSC2U J003536.0+015314 &  00 35 35.97 &  +01 53 14.0  & -$139.0\pm1.5$  & -$361.2\pm1.3$ & 1991.76  \\
GSC2U J010456.6+212002 &  01 04 56.62 &  +21 20 02.0  & -$208.4\pm2.9$  & -$434.1\pm2.8$ &  1991.63  \\
GSC2U J010458.1+212021 &  01 04 58.10 &  +21 20 21.0  & -$205.7\pm2.8$  & -$437.4\pm2.5$ &  1991.63  \\
GSC2U J025936.9+172533&  02 59 35.86 &  +17 25 33.0  &-$271.4\pm8.5^{a}$ &-$93.5\pm7.8^a$&1991.79\\
GSC2U J030339.2+140805 &  03 03 39.15 & +14 08 05.0 & $50.8\pm1.9^{b}$ & -$477.4\pm35.5^b$ &1991.79\\
GSC2U J031000.4+163022 &  03 10 00.43 &  +16 30 22.0  & $208.5\pm4.1^{a}$& -$193.3\pm4.4^a$&1991.79 \\
GSC2U J071858.8+551406  &  07 18 58.80 &  +55 14 06.0  & $161.7\pm3.9$  & -$191.6\pm7.8$ & 1989.93 \\
GSC2U J113524.7+271741  &  11 35 24.72 &  +27 17 41.4  & -$349.4\pm4.0$ &-$62.5\pm3.8$  &1995.31\\
GSC2U J114625.7$-$013635    &  11 46 25.68 &  -01 36 34.8  &$351.8\pm0.6$  &-$433.7\pm1.4$ & 1995.00\\
GSC2U J122810.3+415003   &  12 28 10.32 &  +41 50 03.5  & -$262.9\pm4.5$  & $64.8\pm5.2$ &1988.22\\
GSC2U J123752.6+415622  &  12 37 52.56 &  +41 56 22.4  &    -$337.2\pm2.8$ &$340.6\pm0.7$ &1988.22       \\
GSC2U J123915.4+452520  &  12 39 15.36 &  +45 25 20.4  &   -$637.1\pm1.7$      &  $-530.5\pm3.1$&  1988.22 \\
GSC2U J124212.6+295801   &  12 42 12.57 &  +29 58 00.6  & -$515.1\pm4.2$  &-$350.9\pm19.8$ &1990,08\\
GSC2U J125313.0+343712  &  12 53 12.96 &  +34 37 11.7  & -$303.9\pm7.6$  &$161.2\pm2.1$& 1990.21 \\
GSC2U J125509.6+465520  &  12 55 09.60 &  +46 55 19.9  & -$1087.9\pm2.7$ & $-84.2\pm3.8$&1988.22  \\
GSC2U J125946.8+273404  &  12 59 46.80 &  +27 34 03.6  & -$311.5\pm 10.4$  &$81.3\pm8.6$ &1993.29   \\
GSC2U J131147.2+292348  &  13 11 47.28 &  +29 23 47.4  & -$384.9\pm 2.5^{c}$   &$320.3\pm4.2^c $&  1995.23  \\
GSC2U J131749.0+215715 &  13 17 48.98 &  +21 57 14.7 & -$26.9\pm 1.9$    &-$274.8\pm1.6$ & 1996.30\\
GSC2U J132325.2+303615  &  13 23 25.20 &  +30 36 15.3  & -$101.6\pm2.4$    &-$405.2\pm5.3$ &1995.31\\
GSC2U J132857.4+445034  &  13 28 57.36 &  +44 50 34.5  &$106.4\pm2.7$    &-$223.6\pm3.1$ & 1994.42 \\
GSC2U J132909.6+310947  &  13 29 09.60 &  +31 09 47.2  & -$244.8\pm3.7$    &  $89.1\pm0.7$ & 1995.31  \\
GSC2U J133059.8+302955  &  13 30 59.76 &  +30 29 55.1   & -$471.9\pm2.1$ &  -$107.5\pm1.8$  & 1995.31  \\
GSC2U J133250.9+011707  &  13 32 50.88 &  +01 17 07.2 &$13.4\pm1.6$ & -$274.6\pm2.1$ &1996.23\\
GSC2U J133360.0+001654    &  13 33 60.00 &  +00 16 54.3 & -$328.9\pm2.0$   & $206.1\pm2.0$ & 1996.22  \\
GSC2U J133616.1+001733 &  13 36 16.08 &  +00 17 33.0  & -$280.0\pm0.9$ &  -$139.8\pm1.3$  &1996.23 \\
GSC2U J133620.4+364829   &  13 36 20.40 &  +36 48 29.2  & -$248.4\pm5.3$    &$222.5\pm4.1$ &  1996.22\\
GSC2U J134043.4+020348      & 13 40 43.44 &  +02 03 48.2 & -$531.3\pm2.3$ &   $31.5\pm0.9$ & 1996.23 \\
GSC2U J134107.9+415004   &  13 41 07.92 &  +41 50 04.5 & -$217.0\pm1.4$ &  -$188.7\pm3.3$  & 1996.22 \\
GSC2U J135118.4+425317   &  13 51 18.35 &  +42 53 16.8  &$213.7\pm2.8$   &-$253.8\pm3.8$  & 1994.42 \\
GSC2U J155611.5+115351  &  15 56 11.52 &  +11 53 51.4  & -$193.1\pm3.7$ &   $26.0\pm6.0$ &  1994.42\\
GSC2U J155721.8+141212      &  15 57 21.75 &  +14 12 11.9  & -$204.7\pm1.4$ &-$244.2\pm4.8$   & 1994.44 \\
GSC2U J162242.0+670112  &  16 22 42.00 &  +67 01 12.3  & -$76.2\pm3.6$  & $181.5\pm2.8$ &  1994.42  \\
GSC2U J165237.4$-$011354     &  16 52 37.44 &  -01 13 54.5  &$131.3\pm1.7$ &-$359.1\pm0.9$  &   1990.40 \\
GSC2U J222233.1+122140  &  22 22 33.12 &  +12 21 40.3  &$706.3\pm2.3$ &$185.5\pm2.4$ &     1987.72\\
GSC2U J224011.2$-$030316 &  22 40 11.25 &  -03 03 16.0   & -$202.1\pm5.3$ &-$217.7\pm2.3$  & 1995.59    \\
GSC2U J231518.5$-$020942   &  23 15 18.48 &  -02 09 41.98  &    $578.4\pm2.1$ & $169.5\pm1.3$&1990.66 \\
GSC2U J232115.4+010214  &  23 21 15.36 &  +01 02 14.35   & -$104.6\pm2.4$ & -$244.8\pm1.9$ &1990.66 \\
GSC2U J232116.8+010227 &  23 21 15.84 &  +01 02 26.99   & -$98.0\pm6.6$ & -$242.5\pm3.3$ &1990.66 \\
GSC2U J233539.1+123048  &  23 35 39.12 &  +12 30 48.24     &-$111.3\pm1.5$ & -$134.1\pm1.9$ & 1995.82\\
GSC2U J233707.4+003240 &  23 37 07.44 &  +00 32 40.33   &$299.7\pm1.9$ &$158.5\pm1.6$ &  1990.74\\
\hline
\end{tabular}

\begin{minipage}{16cm}

 $^{(a)}$ Proper motions computed from POSS-I, SERC-J, SERC-ER, and SERC-I positions
 $(\alpha, \delta)$ extracted from the GSC-II database.\\
 $^{(b)}$ Proper motions computed from POSS-II positions $(\alpha,
\delta)$ extracted from the GSC-II database $(\Delta T=6.0$
years).\\
 $^{(c)}$ The value ($\mu_{\delta}=0.286''~$yr$^{-1}$) reported
in Carollo et al. (2002) was affected by a systematic error.
We report here the updated value.\\

\end{minipage}
\end{table*}

\begin{table*}

\caption{The sample of cool WDs from the GSC-II survey:
photometry, classification, source telescope, WD ID in other
catalogues.}

\centering                          
\begin{tabular}{llcccccr}        

\hline\hline
 ID         & &$B_{J}$ & $R_{F}$ & $B_{J}-R_{F}$ & Type & Source & WDs ID\\

\hline


GSC2U J003209.6$-$025402  & &17.78& 16.51  &  1.27  &   DA  & $^a$  &  WD 0029-032\\
GSC2U J003536.0+015314  & &15.78  & 15.70   &  0.08  &   DA  & TNG & WD 0033+016\\
GSC2U J010456.6+212002  & &18.60   & 17.42  &  1.18 &  DC  & TNG & WD 0102+210A\\
GSC2U J010458.1+212021  & &18.90 & 17.52  &  1.38  & DC  & TNG& WD 0102+210B\\
GSC2U J025936.9+172533  & &19.61 & 19.06  &  0.55  &  DA  & TNG&\\
GSC2U J030339.2+140805  & & 20.63  & 19.54 & 1.09   &  DA+dM & TNG&\\
GSC2U J031000.4+163022  & &18.04 & 17.85  &  0.19  &  DA  & TNG&\\
GSC2U J071858.8+551406  & &20.84 & 19.71  & 1.13  & DC  &WHT &\\
GSC2U J113524.7+271741  & & 19.13  & 18.22  &  0.91 &  DC &TNG&\\
GSC2U J114625.7-013635  & & 16.53 & 15.98 &  0.55  &  DA &TNG& SDSS J114625.77-013636.9\\
GSC2U J122810.3+415003  & &19.24& 18.42  &  0.82  &  DC & TNG&\\
GSC2U J123752.6+415622  & &17.83& 16.70  &  1.13  &  DQ & TNG/INT&\\
GSC2U J123915.4+452520  & &16.83& 16.27  &  0.56  &  DA & TNG & WD 1236+457\\
GSC2U J124212.6+295801  & &18.34& 17.75  &  0.58  &  DA & TNG&\\
GSC2U J125313.0+343712  & &19.90& 18.41  &  1.49 &  DC  &TNG&\\
GSC2U J125509.6+465520  & &19.49& 18.28  &  1.21 & DC   &TNG&\\
GSC2U J125946.8+273404  & & 15.55& 15.13  &  0.42 & DA & TNG& WD 1257+278\\
GSC2U J131147.2+292348  & &19.56& 18.04  &  1.52 &  Peculiar DQ$^{b}$ &WHT&\\
GSC2U J131749.0+215715  & & 17.30&16.59     &  0.71 &  DC  &TNG &\\
GSC2U J132325.2+303615  & & 19.89& 19.35  &  0.54 &  DA  &TNG &\\
GSC2U J132857.4+445034  & & 18.57& 18.04  &  0.53 &  DC  &TNG &\\
GSC2U J132909.6+310947  & & 21.23&  19.32 &  1.91 &  DC  &TNG/WHT &\\
GSC2U J133059.8+302955  & & 16.62& 15.75  &  0.87 &  Peculiar DZ$^{c}$ &TNG & WD 1328+307 \\
GSC2U J133250.9+011707  & & 17.34 & 16.65 &  0.69 & DC+DA$^d$ & TNG  &WD 1330+015 (A+B) \\
GSC2U J133360.0+001654  & & 19.50& 17.98  &  1.52 & DQ  &TNG/INT &WD1331+005\\
GSC2U J133616.1+001733  & & 17.72& 16.73  &  0.99  &  DA+dM &TNG&SDSS J133616.05+001732.7\\
GSC2U J133620.4+364829  & &18.21& 17.40  &  0.81   &  DA  &TNG &\\
GSC2U J134043.4+020348  & & 18.38& 17.60  &  0.78  & DA   &TNG &SDSS J134043.35+020348.3\\
GSC2U J134107.9+415004  & & 17.95& 17.52  &  0.43 &  DA   &TNG &\\
GSC2U J135118.4+425317  & &17.14& 16.52 & 0.62  &  DC   &TNG &\\
GSC2U J155611.5+115351  & &  20.51&18.94  &  1.57 &  DC   &TNG &\\
GSC2U J155721.8+141212  & & 19.43& 18.06  &  1.37  &  DC &TNG  &\\
GSC2U J162242.0+670112  & & 18.89& 18.99  & -0.10 &  DA   & TNG/WHT &\\
GSC2U J165237.4$-$011354  & & 17.87& 16.69   &  1.18 &    DA &TNG  &\\
GSC2U J222233.1+122140  & & 19.74& 17.87  &  1.87  & DC  & TNG& SDSS J222233.90+122143.0\\
GSC2U J224011.2$-$030316  & & 18.95&17.79  &  1.16  &   DC  &TNG &\\
GSC2U J231518.5$-$020942  & & 16.39& 16.00  &  0.39  &  DZ &  TNG&WD 2312-024\\
GSC2U J232115.4+010214  & &19.62& 18.65  &  0.97   & DA  &TNG &WD 2318+007A\\
GSC2U J232116.8+010227  & &20.23& 18.79  &  1.44   & DC &TNG &WD 2318+007B\\
GSC2U J233539.1+123048  & &18.81& 18.45  &  0.36  &  DC & TNG &\\
GSC2U J233707.4+003240  & &18.58& 17.61  &  0.97  & DA  &TNG &SDSS J233707.68+003242.3\\
\hline

\end{tabular}

\begin{minipage}{16cm}
$^{a}$ Classification from Monet et al. (1992) and Bergeron et al. (1997)\\
$^{b}$ Peculiar DQ with extremely strong C$_2$ bands (Carollo et al. 2002, 2003)\\
$^{c}$ Heavily blanketed DZ (G 165-7) (Wehrse \& Liebert 1980) \\
$^{d}$ Unresolved binary (DC + magnetic DA) modeled by Bergeron et
al. (1997)
\end{minipage}


\end{table*}

\begin{table*}

\caption{The sample of cool WDs from the GSC-II survey:
cross-match with Luyten's catalogues (LHS and NLTT). Photometry
from SDSS and 2MASS.}

\centering                          
\begin{tabular}{llcccccccc}        

\hline\hline

            & LHS/NLTT & \multicolumn{5}{c}{SDSS} &  \multicolumn{3}{c}{2MASS} \\
 ID         &          & u & g    & r    & i    & z & J      & H     & K   \\

 \hline
GSC2U J003209.6$-$025402& LHS 1093 & -  & - & - & - &-  &15.56 & 15.37 & 15.35   \\
GSC2U J003536.0+015314 &LP 585- 53 & -  & - & - & - & - &15.65 & 15.52 & 16.12\\
GSC2U J010456.6+212002 & LHS 5023 & - & - & -  &-  &-  & 16.52& 16.50 & 15.55   \\ 
GSC2U J010458.1+212021 & LHS 5024 & - & - & -  &-  &-  & 16.73& 16.27 & 15.59   \\ 
GSC2U J025936.9+172533 &LP 411- 34 & - & - & -  & - &-  & -    &   -   &  -   \\
GSC2U J030339.2+140805 & -          & - & - & -  & - &-  &14.01 & 13.56 & 13.25 \\
GSC2U J031000.4+163022 & LP 412- 7 & - & - & -  &  -&-  &  -   &-      &-  \\
GSC2U J071858.8+551406 & -          & - & - & -  & - & - &  -   &-      &- \\
GSC2U J113524.7+271741 & LP 318-432 & - & - & -  & - & - &  -   &-      &- \\
GSC2U J114625.7$-$013635&LHS 2455 &17.14 & 16.51 &16.24 &16.14 &16.17&15.54 & 15.38 & 15.18  \\ 
GSC2U J122810.3+415003 &LP 217- 20 & 19.89 &19.02 &18.57 &18.41&18.29 &  - &-&-  \\
GSC2U J123752.6+415622 & LHS 5222 &17.85 &17.75 & 17.11& 16.85& 16.93 & 16.55 & 16.50 & 15.48  \\ 
GSC2U J123915.4+452520 & LHS 2596 &17.35 &16.65 &16.37& 16.28 & 16.29  & 15.60 & 15.20 & 15.73  \\ 
GSC2U J124212.6+295801 & LHS 2607 &-      &-       & -    & -     & - & -     &  -    &  -     \\ 
GSC2U J125313.0+343712 & -          & -     & -      & -    & -     &  -     &  -    &-      &- \\
GSC2U J125509.6+465520  & -          & 21.05 &19.17  &18.38 & 18.05  & 17.92     & -     &-      & -  \\
GSC2U J125946.8+273404 & LP 322- 267&-     &-       & -    & -     & -      &  15.13&  14.98&  14.99 \\
GSC2U J131147.2+292348 & -           & 19.70&19.45   & 18.50&17.94  &18.00   &  17.48&  17.13&  17.08\\
GSC2U J131749.0+215715 & LP 378- 956&-     &-       &   -  & -     & -      & 16.03 & 15.79 & 15.96  \\
GSC2U J132325.2+303615 & -           & -    &-       &-     & -     & -      & -     & -     & -    \\
GSC2U J132857.4+445034 & -           & 18.79&18.41   &18.33 &18.34  & 18.40  & -     &-      &-   \\
GSC2U J132909.6+310947 & -           & -    & -      & -    & -     & -      & -     &-      &-    \\
GSC2U J133059.8+302955 &LHS 2745     & 18.27 &16.28  &15.87 &15.90  & 16.08  & 15.40 & 15.28 & 15.41    \\
GSC2U J133250.9+011707 & LP 618-1    &17.72  &17.13  &16.95&16.91   &16.92   & 16.40 & 16.30 & 15.80    \\
GSC2U J133360.0+001654 & LP 618- 6   & 19.07&19.40   &18.33 & 18.07 & 18.14  & -     &-      &-   \\
GSC2U J133616.1+001733 & LP 618- 14  &17.97 &17.41 & 17.05&16.29  &15.66 &14.26  & 13.74 & 13.51 \\ 
GSC2U J133620.4+364829 & LP 269- 25  &18.87 & 18.04  & 17.66   &17.51 &17.48 & -     &-      &-    \\
GSC2U J134043.4+020348 & LHS 2781 & 19.26&18.11   &17.62 &17.43     & 17.36   & -     &-      &-   \\ 
GSC2U J134107.9+415004 & LP 219- 53  & 18.06&17.67   &17.60 &17.61    & 17.69      & -     &-      &-    \\
GSC2U J135118.4+425317 & LP 219- 80  &17.62     &17.05       &16.87 &16.86  &16.93 & 16.38 & 15.92 & 17.23   \\
GSC2U J155611.5+115351 & -           & -    & -      & -    & -     & -      & -     &-      &-   \\
GSC2U J155721.8+141212 & LP 504- 10  & -    & -      & -    & -     & -      & -     &-      &-   \\
GSC2U J162242.0+670112 & -           & -    & -      & -    & -     & -      & -     &-      &- \\
GSC2U J165237.4$-$011354 & LP 626- 11 &-     &-       & -    & -     & -      & 16.19 & 15.96 & 17.0 \\
GSC2U J222233.1+122140 & -       &19.48 & 17.90  & 17.30& 17.07 &16.97       &-      &-      &- \\ 
GSC2U J224011.2$-$030316 & LP 700- 65& - & -& -      &-     &-      &  16.93 & 16.56 & 17.02    \\
GSC2U J231518.5$-$020942 & LHS 3917   & - & -&-       &-     & -     & 15.49  & 15.75 & 14.86    \\
GSC2U J232115.4+010214 & LP 642- 52 &20.51&19.37&18.84&18.65 & 18.59& -&-&-  \\
GSC2U J232116.8+010227 & LP 642- 53 &21.83&19.85&18.94& 18.65 & 18.47 & -&-&-  \\
GSC2U J233539.1+123048 & -  & - & - & - & - & - & -&-&- \\
GSC2U J233707.4+003240 & LP 643- 20&19.34&18.28&17.79&17.61&17.60& 16.55& 16.17 & 15.35 \\ 
\hline

\end{tabular}

\end{table*}

\begin{figure*}

\resizebox{\hsize}{!}{\includegraphics{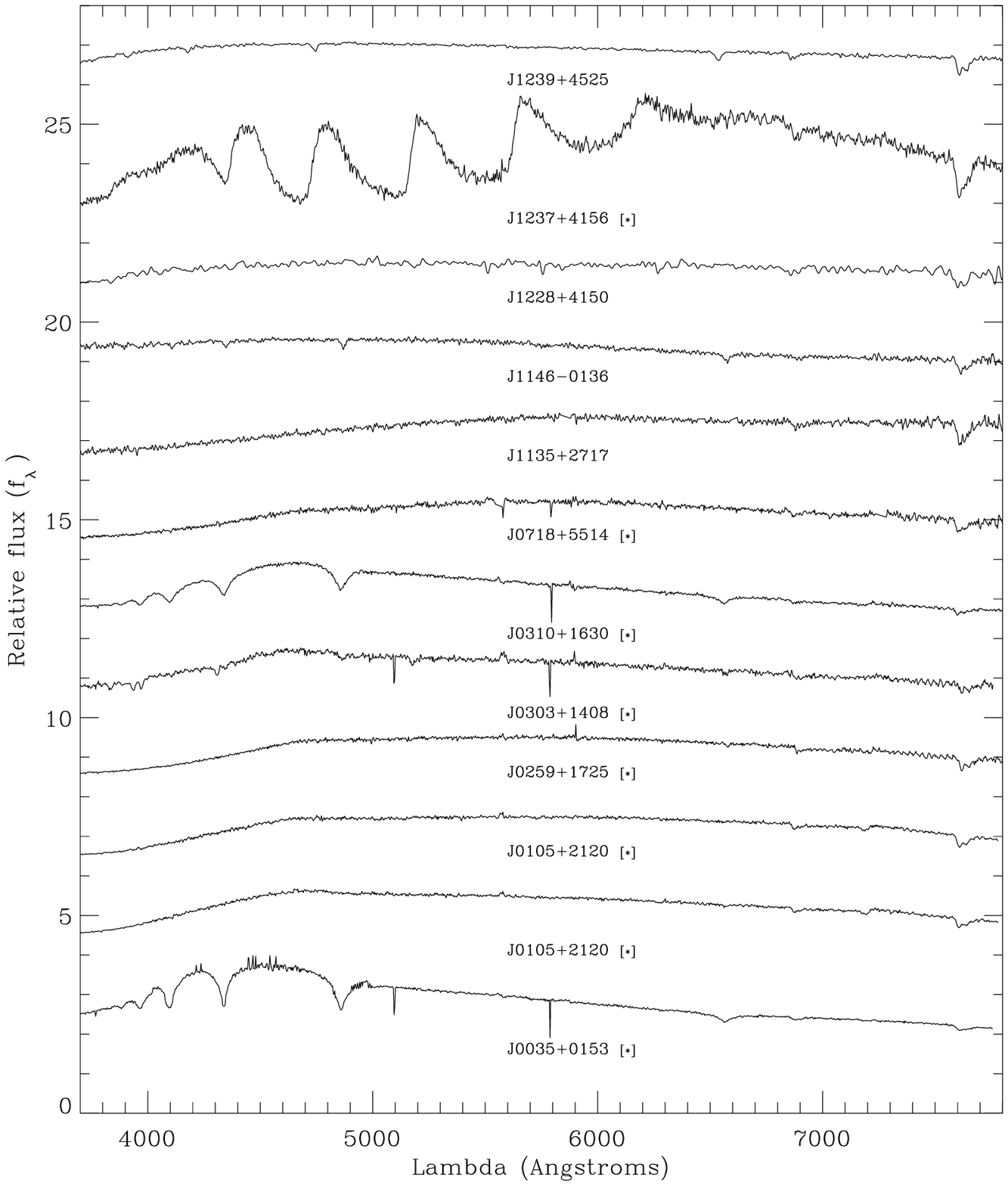}}

      \caption{Optical spectra. All the spectra are
               normalized at 5500 \AA, shifted vertically
               from each other by arbitrary units, and ordered
               by increasing right ascension. Asterisks
               indicate spectra not fully calibrated, shown for
               classification only.
               }

         \label{Fig3}
\end{figure*}

\begin{figure*}
\resizebox{\hsize}{!}{\includegraphics{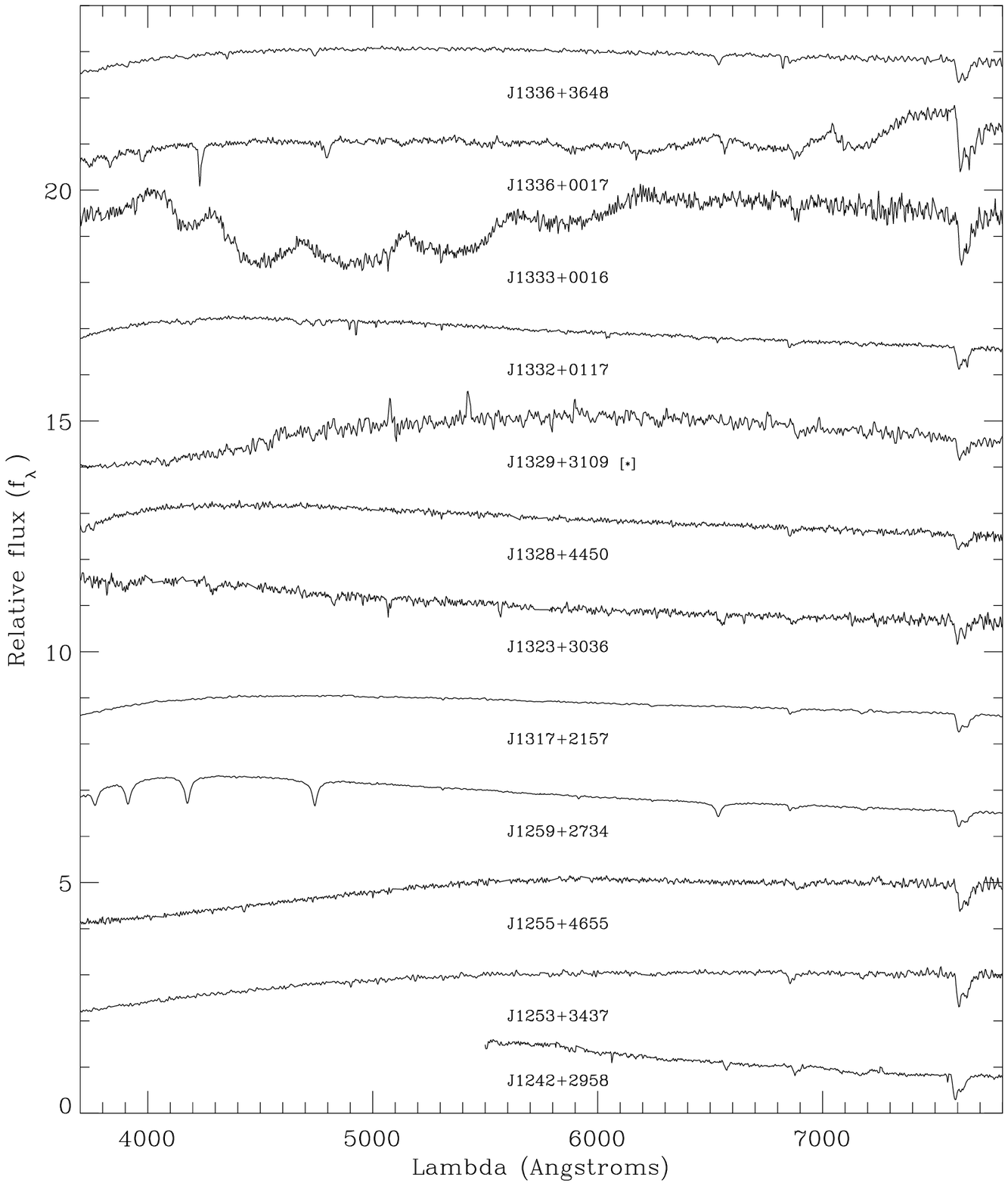}}

      \caption{Optical spectra. All the spectra are
               normalized at 5500 \AA, shifted vertically
               from each other by arbitrary units, and ordered
               by increasing right ascension. Asterisks
               indicate spectra not fully calibrated, shown for
               classification only.
                             }
         \label{Fig4}
\end{figure*}

\begin{figure*}
\resizebox{\hsize}{!}{\includegraphics{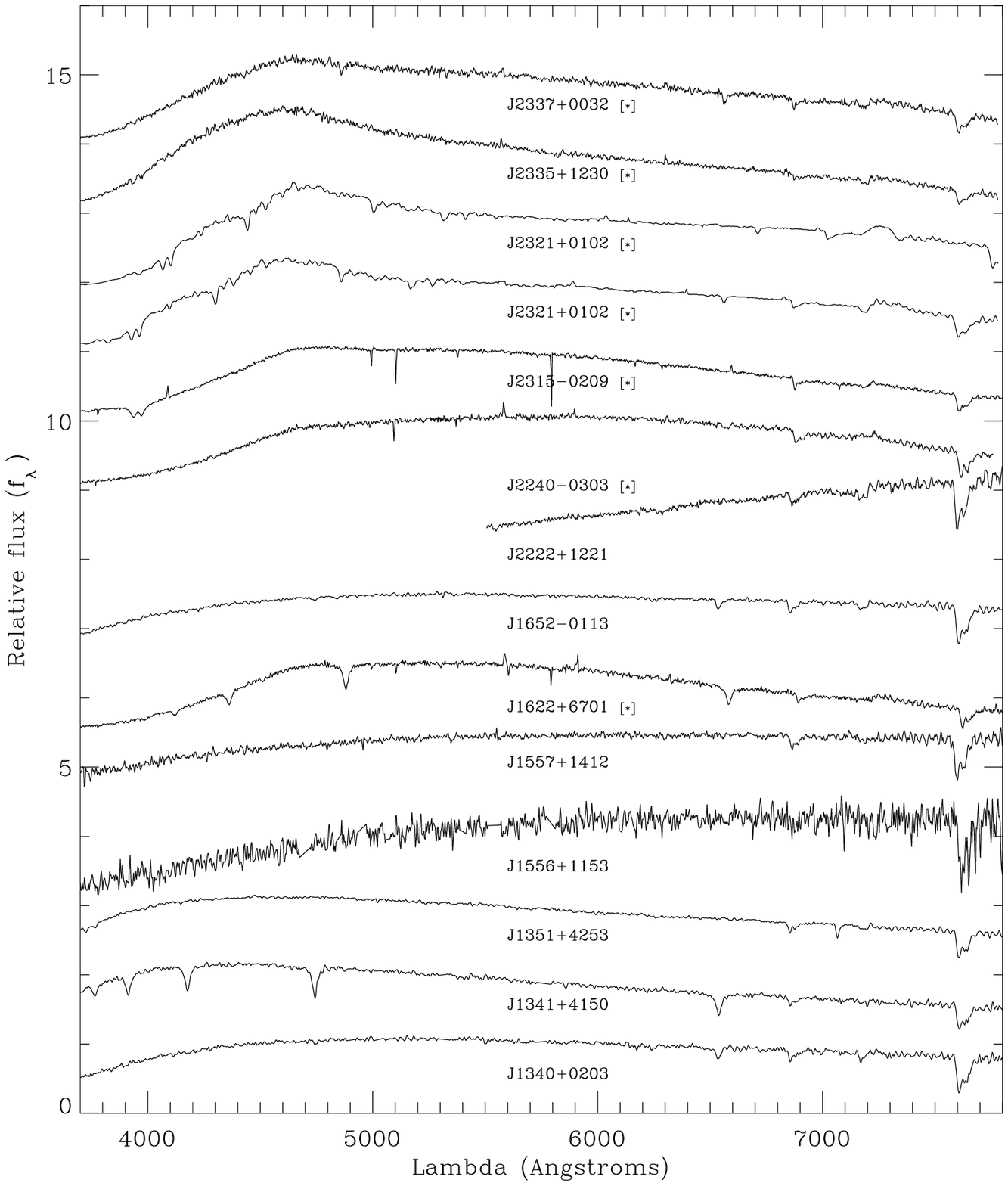}}

      \caption{Optical spectra. All the spectra are
               normalized at 5500 \AA, shifted vertically
               from each other by arbitrary units, and ordered
               by increasing right ascension. Asterisks
               indicate spectra not fully calibrated, shown for
               classification only.
                             }
         \label{Fig4}
\end{figure*}





\begin{figure*}
\resizebox{\hsize}{!}{\includegraphics{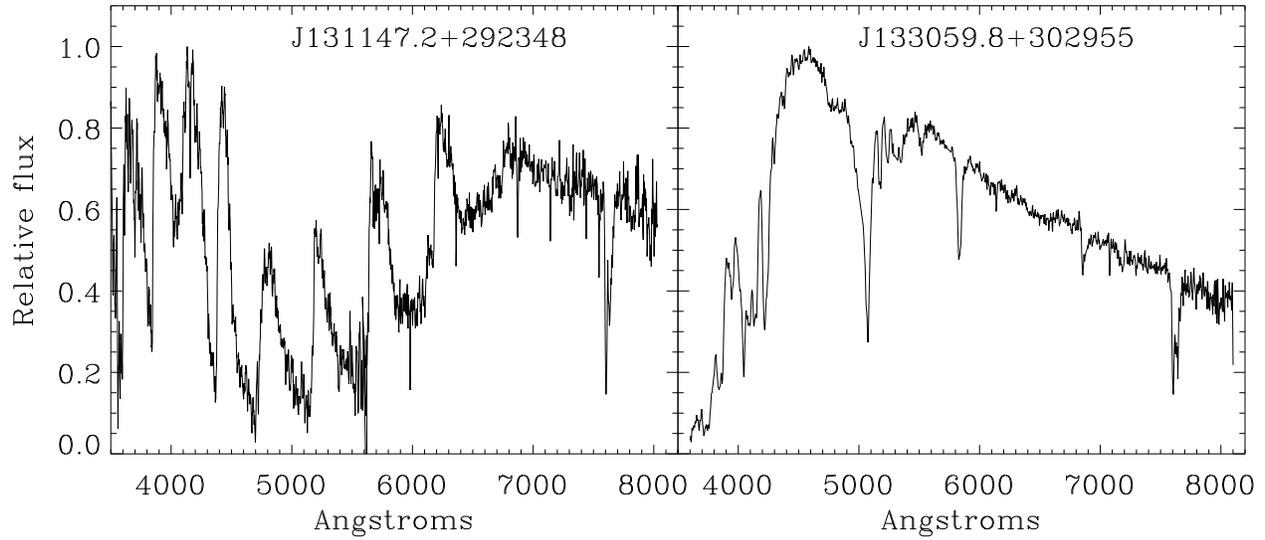}}

      \caption{Optical spectra of the peculiar WDs. Left
      panel shows GSC2U J 131147.2+292348 which is the
      coolest DQ with extremely strong C$_2$ bands. Right panel
      shows WD 1328+307 (G 165-7) which is a heavily blanketed DZ.
                      }
         \label{Fig5}
\end{figure*}






\end{document}